\documentclass[graybox, nosecnum]{svmult}

\usepackage{mathptmx}       
\usepackage{helvet}         
\usepackage{courier}        
\usepackage{type1cm}        
\usepackage{makeidx}         
\usepackage{graphicx}        
\usepackage{multicol}        
\usepackage[bottom]{footmisc}
\usepackage{hyperref}        
\usepackage{soul}            
\hypersetup{colorlinks=true,urlcolor=blue}
\usepackage[square,numbers]{natbib}

\usepackage{amsmath}
\usepackage{amssymb}

\usepackage{acronym}

\makeindex             

\begin{document}
\title*{The Major Gamma-ray Imaging Cherenkov Telescopes (MAGIC)}
\author{O. Blanch \thanks{corresponding author} and J.Sitarek \thanks{corresponding author} }
\institute{O. Blanch \at Inistitut de F\'{i}sica d'Altes Energies (IFAE) - The Barcelona Institute of Science and Technology (BIST), E-08193 Bellaterra (Barcelona) Spain, \email{blanch@ifae.es}
\and J. Sitarek \at University of Lodz - Faculty of Physics and Applied Informatics - Department of Astrophysics, 90-236 Lodz Poland, \email{jsitarek@uni.lodz.pl}}
%
%
\maketitle


%
\abstract{
The MAGIC telescopes, located at  Observatorio El Roque de los Muchachos (La Palma, Spain) are two Imaging Air Cherenkov Telescopes observing the Very High Energy (VHE) gamma rays. They are run by an international collaboration composed of over 40 institutions from 12 countries. The first telescope was inaugurated in October 2003. The commissioning of the second finished in 2008. The MAGIC telescopes were designed to lower the energies to which ground based telescopes had access as well as to be able to point to any direction in the sky in less than 25 seconds. The former required the large reflective surface of 17 meters as well as an effort to optimise the mirror reflectivity and photo sensor sensitivity. The latter was achieved by minimising the weight of the full instrument using for instance carbon fibre reinforced plastic tubes for the mirror frame. The sensitivity of the MAGIC telescopes have been improving over the years thanks to hardware upgrades as well as new analysis techniques, which allowed the collaboration to keep a rich scientific program. The discovery of VHE emission from Gamma Ray Bursts and pulsars have called for a revision of the models that explain the production of gamma rays there. Both the observation of sources in flaring state as well as a systematic monitoring of sources have provided valuable data to better understand astrophysical sources both in our Galaxy and outside it.  Relevant constraints on fundamental quantities like dark matter cross-section, quantum gravity scale and density of extragalactic background light have also been extracted from the observations.
}

\section{Keywords} 
MAGIC Telescopes, Gamma-rays, Very-High-Energy, Astroparticle Physics, IACT, Cherenkov Telescopes
\section{Introduction}

Gamma rays are the most energetic form of electromagnetic radiation. The individual photons have energies above 100 keV and they are absorbed by the Earth atmosphere. This makes it difficult to observe them with ground based instruments. Hence, gamma ray astronomy only started to develop when it became possible to get detectors high in or above the atmosphere, which happened around 1960. Alternatively, gamma rays can also be indirectly detected with ground-based instruments by looking at the products of their interaction with the atmosphere. In particular, Imaging Atmospheric Cherenkov Telescopes (IACTs) have been proven to be very successful. 

The Whipple 10m telescope \citep{Kildea:2007zz} was the first IACT to detect gamma rays from an astrophysical source. The IACT technique is based on the detection of Cherenkov photons that are produced in the extensive air showers induced by high energy particles entering the atmosphere. Those Cherenkov photons reach the ground over a circle larger than 100 m diameter with a quite uniform density. Hence, a single telescope will be able to collect the secondary Cherenkov photons in an area $\sim 10^{5}$ $m^{2}$, which is much larger than physical size of the detector (in particular comparing to the gamma-ray detectors high up in or above the atmosphere). The size of the mirror provides the amount of Cherenkov photons that are collected for each incident particle. The larger the mirror surface the lower energy is accessible. 
The key characteristic that distinguishes IACTs comparing to earlier waveform sampling approaches is that IACTs are not only able to register the amount of Cherenkov light reaching the telescope, but also provide an image of each shower developing in the atmosphere. 
This turned out to be essential in the efficient reconstruction of the events and rejection of Cosmic-Ray-induced background. 
In general, IACTs are sensitive to gamma-rays with energies between 100 GeV and 100 TeV, the Very High Energy (VHE) domain. MAGIC (Major Atmospheric Gamma Imaging Cherenkov) is an instrument belonging to the third generation of IACTs (i.e. small arrays of medium/large IACTs constructed in 2000s).  

Gamma rays are produced in the most energetic astrophysical sources and many of them could emit gamma rays reaching the energy domain of the IACTs. The VHE gamma-ray emitter can be a Galactic object such as a pulsar, supernova remnant, microquasar, gamma-ray binary or a nova. But they can also have an extragalactic origin with several types of Active Galactic Nuclei being already established as emitters. The the VHE sky is not only allow us to probe the most energetic phenomena but it is also changing very fast. Many of the astrophysical sources show variability at VHE with different time scales and there are event sources like the Gamma Ray Burst shining only for some seconds and then fading away for ever. 
A more complete introduction on gamma-ray astronomy and IACTs is available at \cite{universe8040219}.





\section{The MAGIC history}


In this section we introduce the history of MAGIC: starting from the idea behind it, through the start of operations and evolution of the MAGIC observational program.

\subsection{The MAGIC Collaboration}

For a long time there had existed a gap between those energy to which ground and space based gamma-ray detectors were sensitive. In 1994,  the first ideas to build a Cherenkov Telescope able to close the gap started to crystallise and the history of MAGIC started. Also at that time, E. Lorenz started to look for institutions that wanted to join the adventure. In 1996, there was already a critical mass that would become the MAGIC collaboration. Still, the formal collaboration only started in 1998 with the goal to build the first MAGIC telescope (MAGIC-I). An updated Memorandum of Understanding already including the second MAGIC telescope (MAGIC-II) was approved in 2005. As of today, the MAGIC collaboration is still active. It was in charge of building and commissioning the two MAGIC Telescopes that are both dedicated to the memory of Florian Goebel, being its full name the MAGIC Florian Goebel Telescopes. Florian Goebel was a brilliant, warm and motivating scientist who easily assembled successful teams around him. An unfortunate accident took his life while he was working on MAGIC-II, just a few days before its planned inauguration in 2008.  Florian has become an inspiration for every new generation of physicists in the MAGIC collaboration. The telescopes are surrounded by several auxiliary instruments like an elastic LIDAR \citep{MAGIC-LIDAR} and its data centre is hosted at PIC (Port d'Informaci\'{o} Cient\'{i}fica). The MAGIC Collaboration keeps maintaining, operating and scientifically exploiting all MAGIC related instrumentation. 

Currently, the MAGIC collaboration is composed of over 40 institutions from 12 countries (Armenia, Brazil, Bulgaria, Croatia, Finland, Germany, India, Italy, Japan, Poland, Spain and Switzerland). The individual members account to more than 250 people being most of them astrophysicists but also including engineers and research support personnel. In addition, there are currently about 50 associated scientists. The collaboration is organised through two bodies: the Collaboration Board for the financial and strategical decision and the Executive Board for planning operation, maintenance and scientific exploitation. The latter is structured in the different Physics Working Groups (Galactic, AGN, Fundamental Physics and Transient) as well as service bodies (Software, Hardware, Outreach, ...). MAGIC is an equal opportunity collaboration committed to diversity in the workplace and sensible to the worries of people in their early career stages, for which a specific committees exist. To recognise relevant contributions from young members of the collaboration the Florian Goebel Prize is awarded by the Collaboration Board of MAGIC to young members that contributed to important achievements.

\subsection{Envisioned Scientific Goals}

Back in 1996, the most detailed gamma-ray data had been provided by the EGRET satellite \citep{EGRET} at the lowest energies ($\sim$ GeV) and  the Cherenkov telescopes Whipple \citep{Kildea:2007zz} and HEGRA \citep{Daum:1997fp} at the highest ones ($\sim$ TeV). The latter had been able to detect only a handful of sources due to their limited sensitivity. In this scenario, the construction of MAGIC was aiming at both closing the energy gap between them from $\sim 10$~GeV to $\sim 250$~GeV   as well as to improve the sensitivity at TeV energies.
A Cherenkov telescope with the lowest possible threshold would succeed in both, the latter thanks to large zenith observations. Also at the same time, the next generation of satellite-borne instruments were being built (on board of {\it AGILE} \citep{Tavani:2008zz} and {\it Fermi} \citep{Fermi-LAT:2012fsm}). They would also aim to cover the unexplored window. Due to the difference in collection area, the ground based telescopes were expected to be better suited for fast variable emissions in this energy range. 

The prime example of fast phenomena emitting a huge amount of energy are the Gamma Ray Bursts (GRBs). The characteristic time of the GRBs ranges from a few seconds to a few hundreds of seconds and it was not only one of the main scientific goals of MAGIC but also a design driver. The full MAGIC telescopes were designed to be able to move fast enough to catch the GRBs. The large collection area at low energies was also expected to be instrumental to study Active Galactic Nuclei (AGNs) because they show periods of flaring activity with fast changes on the gamma-ray emission. In addition, previous ground based gamma-ray detectors could only detect gamma-ray emitters in the very local universe (z $\sim$ 0.1) since for more distant sources the gamma rays were absorbed while travelling from their origin to the Earth. 

The low energy threshold was also expected to allow studies of pulsar and stellar accretion driven systems. The improved sensitivity at TeV energies would allow to test the Supernova Remnants as the sources of Galactic Cosmic Rays.

The improved sensitivity of MAGIC was also important to check for fundamental physics questions. The energies in place in the astrophysical objects and the distances travelled by the photons emitted there give insights to energies and phenomena that cannot be explored in the human made accelerators. The search for dark matter or Lorentz Invariance Violation 
as well as measurements of cosmological variables 
or cosmological magnetic fields  were in the physics program from the beginning.

\subsection{First Light and start of MAGIC-I operation}

\begin{figure}
    \centering
    \includegraphics[width=0.9\textwidth]{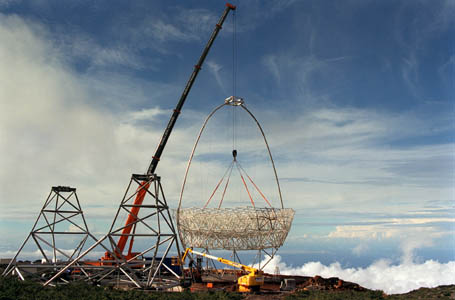}
    \caption{Mirror and camera support structure of the MAGIC-I Telescope being positioned on top of the azimuth rail. Credit: R. Wagner.}
    \label{MAGIC1-StrucureInstallation}
\end{figure}

The dish structure of the first MAGIC telescope was placed on top of the foundations on 14 December 2001 (Figure~\ref{MAGIC1-StrucureInstallation}). The reflective surface and the camera were still missing and it was not until October 2003 that MAGIC-I was inaugurated. The telescope was already fully functional despite the mirror surface was not complete and the commissioning of the full system had not finished.

Regular data taking started targeting known TeV gamma-ray emitters, like Crab Nebula, Mrk 421, Mrk 501, 1ES 1426+428 and 1ES 1959+650, that had been already detected by the previous generations of Cherenkov Telescopes. Initially, the observation were restricted to zenith angles below 40 degrees sacrificing the increased collection area at high zenith angles in favour of a lower energy threshold. Moreover, the observations were done in ON-OFF mode. For each source, a similar amount of time was accumulated from a positions in the sky from where no gamma-ray emission was expected. The OFF regions were selected with similar conditions in terms of zenith and azimuth angle as well as of brightness of the surrounding sky. Shortly after the first regular observations, improvements both from the observation and analysis view were done. Studies on the source pointing scheme 
and range 
allowed to optimise the observation strategy. Improvements on the analysis level that allowed to identify better which parts of the image were relevant
and to better differentiate images generated by gamma rays from other 
boosted the sensitivity of the instrument.

\subsection{First scientific results} 

At the start of MAGIC scientific operations in 2003, the Cherenkov astronomy was still a fledgling field with only a dozen of the brightest sources known. 
An important objective at that time was to expand the list of known very-high-energy (VHE) gamma-ray emitters into a full population and to probe for such emission in new classes of objects. 
With its low trigger energy threshold MAGIC gave a unique possibility to study sources that have soft (steep) observed spectra.
The spectra might be intrinsically soft (e.g. due to inefficient acceleration, or strong energy losses of accelerated particles), or become soft due to either internal (inside the source) or external (on the path between the source and the observer) absorption. 

While at that time there was no satellite detector scanning the sky in GeV gamma-rays, the earlier observations of EGRET \cite{1999ApJS..123...79H} left a number of discovered GeV sources, that were a natural target for the search of VHE gamma-ray emission.
This was both the case for associated objects (with a counterpart at e.g. optical or radio wavelengths) as for the unidentified EGRET sources.

The first new sources discovered by MAGIC were mainly blazars, in particular 1ES\,1218+30.4 \cite{2006ApJ...642L.119A} and Mrk\,180 \cite{2006ApJ...648L.105A}.
Moreover, some of the blazars that could have been detected with the previous generation of IACTs only when undergoing a large flare, could be studied with MAGIC also in low states (see e.g. \cite{2007ApJ...662..892A}), providing much more complete information about the behaviour of the source. 
Finally, the previous generation of IACTs was able to study almost solely blazars of the HBL type (High-energy peaked BL Lacs).
In contrast, already in the first phase of MAGIC operations, its low energy threshold allowed detection and studies of sources peaked at lower energies, in particular BL Lacertae (\citealp{2007ApJ...666L..17A}; initially classified as LBL - low energy peaked, later as IBL - intermediate energy peaked) and S5\,0716+714  \cite{2009ApJ...704L.129A}.
The joint radio interferometry and TeV observations of radio galaxy M87 revealed that the emission should occur in vicinity of the black hole \cite{2009Sci...325..444A}.  
All those observations allowed to enlarge the catalogue of VHE gamma-ray emitters, expanding it also with new sub-classes of objects. 
They also provided a valuable insight for cross-band correlations and broadband modelling of blazar Spectral Energy Distributions (SED). 

Moreover, a few important Galactic sources were discovered at the time. 
Among them, the first detection of the variable gamma-ray emission from LS\,I\,+61\,303 \cite{2006Sci...312.1771A} -- the third binary detected in this energy range.
The observations covered 6 orbital periods of the source. The emission was observed in the same phases of the orbit - showing the periodicity of the emission. 
MAGIC was also able to expand the list of known supernova remnants (SNRs), with the detection of IC 443 \cite{2007ApJ...664L..87A}. 

Two most important discoveries in the first years of MAGIC operations, that led to a major breakthrough in astrophysics were discoveries of VHE gamma-ray emission from the Crab Pulsar and from blazar 3C 279.
The Crab pulsar was a known GeV source, however EGRET measurements could not reach high enough energies to explain the mechanism of the emission. 
From theoretical arguments a strong cut-off (exponential or super-exponential) was expected due to internal absorption of the gamma rays in the magnetosphere or in the dense radiation field. 
The low energy threshold of MAGIC made this instrument the perfect choice for studying the Crab pulsar, but the source still provided a major challenge.
In fact a dedicated analogue trigger (Sum Trigger) was constructed and special low-energy optimised analysis procedures were derived. 
This resulted in lowering the energy threshold of the instrument by a factor of two, and consequently the first detection of the Crab pulsar \cite{2008Sci...322.1221A} by an IACT. 
The reconstructed spectrum excluded Polar Gap models, and had a major impact on the development of pulsar models in the next years. 

3C 279 is a blazar of a Flat-Spectrum Radio Quasar (FSRQ) type. 
The detection of VHE gamma-ray emission from it with the MAGIC telescopes \cite{2008Sci...320.1752M} was a major result not only for the physics of FSRQ, but also for cosmology.
FSRQ contain a strong radiation field of a Broad Line Region (BLR) that can efficiently absorb the gamma-ray radiation (see e.g. \citealp{2006ApJ...653.1089L}).
Therefore detection of non-absorbed gamma rays at sub-TeV energies shows that emission must originate beyond the BLR. This provides constraints on the location of the emission region by orders of magnitude better that could be achieved with the angular resolution even at the radio frequencies. 
Second, 3C 279 is located at the redshift of 0.536, thus over a factor of two more distant than VHE gamma-ray emitters known previously.
In fact, at this redshift the absorption in the Extragalactic Background Light (EBL) in the VHE gamma-ray range was thought to be so severe that could possibly prevent the detection of the source. 
Therefore, the detection of VHE gamma-ray emission from 3C 279 by the MAGIC telescopes led to a major reconsideration of the EBL models in the following years. 
A few of the sources detected in the first years (in particular  PG\,1553+113 and  LS\,I\,+61\,303) became regularly monitored by MAGIC in the following years, creating a massive legacy data set. 

\subsection{Going to stereo} 

In 2008 the construction and commissioning of the second, MAGIC-II telescope, was concluded and since 2009 the standard observation mode is joint observations with both telescopes in a stereoscopic mode.
This resulted in a major improvement of performance, especially at the lowest energies, and opened new possibilities for studies. 

The enhanced sensitivity and low-energy performance led to a number of discoveries.
Those features were crucial in further expanding the VHE gamma-ray horizon. 
In 2014 and 2015 two FSRQ objects at the redshift $\sim1$ were detected in VHE gamma rays with the MAGIC observations: QSO\,B0218+357 \cite{2016A&A...595A..98A} and PKS\,1441+25 \cite{2015ApJ...815L..23A}, beating again the record of the most distant source.
This for the first time allowed studies of the VHE gamma ray emission produced at half the current age of the universe. 
Moreover, QSO\,B0218+357 is still the only gravitationally-lensed source with the known delay detected at the VHE gamma rays. 
The delay was used to predict the expected time of the arrival of the VHE gamma-ray emission, leading to the MAGIC detection.  

The excellent performance of the stereoscopic system proved to be helpful in the detection of the new types of VHE gamma-ray sources. 
While the observations of GRBs were in the scientific program since the beginning of the MAGIC concept, it took nearly 15 years, filled with hardware and software optimisations, and following of $\sim 200$ GRB objects, to finally discover the first one.
Due to fast MAGIC reaction, and not too high distance of the source ($z=0.4245$), GRB 190114C showed an enormous signal, reaching in the first seconds the flux level of the order of hundred Crab Nebula units, becoming by far the brightest source ever detected in the VHE gamma rays \cite{2019Natur.575..455M}.
Soon after, the second MAGIC GRB (GRB 201216C) was detected at redshift 1.1, breaking once again the limit of the farthest VHE gamma-ray source \cite{2020ATel14275....1B}.

Similarly, the follow up program of Galactic novae took over ten years until the major discovery of the first VHE gamma-ray nova \cite{2022NatAs...6..689A}. 
RS Oph is a nova of recurrent symbiotic type, and its observations with the MAGIC telescope not only showed that such sources can emit VHE gamma rays, but even more importantly revealed hadronic origin of the emission. 

Also the other MAGIC long-term project: search of coincident gamma-ray and neutrino emission in the quest of studying the hadronic jet models reached a positive outcome with the discovery of VHE gamma-ray emission from a blazar coincident with the observations of a neutrino \cite{2018Sci...361.1378I}.
Somewhat surprisingly however detailed modelling of the gamma-ray emission showed that it is mainly explained by leptonic rather than hadronic processes \cite{2018ApJ...863L..10A}.

Further improvement of the low energy performance of MAGIC was achieved with the construction of the stereoscopic version of Sum Trigger (Sum-Trigger-II, \citealp{2021ITNS...68.1473D}).
It allowed MAGIC detection of a second pulsar (and third VHE gamma-ray pulsar), Geminga \cite{2020A&A...643L..14M}.
Due to the usage of this trigger, and extreme softness of the source the emission could be measured down to 15\,GeV, establishing a record for the ground-based instrument. 

High sensitivity of the stereoscopic system allowed also studies of the emission on shorter time scales. 
Variability of the VHE gamma-ray emission with time scales of (tens) of minutes was observed from some of the studied objects, in particular FSRQ PKS\,1222+21 \cite{2011ApJ...730L...8A}, FSRQ PKS\,1510-089 \cite{2021A&A...648A..23H} and misaligned blazar IC\,310 \cite{2014Sci...346.1080A}.
Such ultra-fast variability is in strong tension with the existing diffuse shock acceleration models in the jets of AGN. 
It stimulated development of alternative theoretical scenarios such as interactions of jets with clouds (see e.g. \citealp{2016MNRAS.463L..26B}) emission originating in the magnetosphere of the black hole (see e.g. \cite{2016ApJ...833..142H}), magnetic field reconnection minijets \citep{2013MNRAS.431..355G}.

As the MAGIC telescopes became a mature instrument, with a good understanding of its performance and systematic uncertainties and accumulated data over the years, deep exposure projects became possible. 
Examples of those involve 253 hrs of observations of the  Perseus Cluster in the search of diffuse gamma-ray emission from Cosmic Ray interactions \cite{2016A&A...589A..33A}, searches of Dark Matter in Dwarf Spheroidal galaxies using  354 hrs \cite{2022PDU....3500912A}, measurement of the Crab pulsar spectrum up to TeV energies with 320 hrs \cite{2016A&A...585A.133A},
EBL studies with a dozen of blazars with 316 hrs of time \cite{2019MNRAS.486.4233A}, 
deep observations of Globular Cluster M15 with 165 hrs \cite{2019MNRAS.484.2876M} 
and a catalogue of Extreme HBL sources covering 265 hrs of data \cite{2020ApJS..247...16A}.

\section{The MAGIC technology}

In this section we introduce the main hardware components of the MAGIC telescopes, and discuss the major upgrades of the system. 


\subsection{The light structure}

The MAGIC Telescopes were born to reach the lowest possible energies. This is very much related to the size of the reflective surface. The larger the surface covered by mirrors is, the lower energies are at reach. The other design driver was the need to be able to change the pointing position fast in order to catch VHE gamma rays from a GRB. The latter implies high drive power and low moving weight. The design of the full MAGIC Telescopes was done keeping in mind the low weight argument. The design was inspired by a 17 m solar concentrator from the ``German Research Institute for Aviation and Space Flight''. Its mirrors and its focal instrumentation were not fitting the need for a Cherenkov Telescope, but the structure was. This fixed the trade off of the largest mirror surface keeping the weight (and price) low and led to an alt-azimuth mount.

The structure can be divided into several parts: the azimuth system, the altitude system, the mirror support and the camera support.

The azimuth system is located directly on top of the foundations and consists of a circular rail, a central axis, six bogeys (chassis with wheels), the undercarriage and the azimuth motors. The undercarriage and the bogeys are steel constructions and need to support the weight of all the other elements. They run on the circular 20 m diameter rail with an angular range of 400 degrees. The altitude system is driven by a single motor and it is coupled to the camera support structure. The total moving weight is 64 tons in azimuth out of which 20 tons is also moving in altitude. Azimuthal movement is achieved by two electric driving motors and the altitude movement by one, each with a maximum output of 11 kW. The camera support is a single aluminium tubular arc, secured against transverse movements by pre-stressed steel cables. Finally, the mirror frame was the most innovative part of the structure. It is made from carbon fibre reinforced plastic tubes, based on a rod-and-knot system without welds by German construction design company MERO. The frame itself weighs less than 20 tons.

\subsection{The mirrors}
 
The reflective surfaces of the MAGIC telescopes have a parabolic shape of 17 meters diameter and the focal length to diameter ratio is f/D=1.03. The actual active reflective mirror surface of each telescope is $\sim$ 240 square meters made of square mirror elements. Originally, MAGIC-I had 974 mirrors of 49.5x49.5 cm diamond-milled aluminium with quartz coating. Four mirror elements were mounted and pre-adjusted on each 1x1m panel.  MAGIC-II consists of 246 99x99 cm mirror panels, the outer part is made up of 104 glass mirrors, the inner part of 142 aluminium panels. Over the years, a sizeable fraction of the MAGIC-I mirrors have been replaced by new aluminium or glass mirrors. The quartz layer is very delicate and can be easily scratched or damaged, which also makes cleaning the mirrors almost impossible. On the other hand, the glass mirrors have as a protective layer an ultra-thin glass sheet which is back-coated with aluminium~\cite{2019ICRC...36..823W}. They can be easily cleaned and almost no degradation in reflectivity is expected.

All types of mirrors installed in the MAGIC telescopes have a light-weight aluminium honeycomb plate of 2-6 cm thickness and are shaped with the needed curvature. The maximum deviation of an individual mirror element from the ideal parabola is less than 10 micrometers and the mean roughness is 4 nm. The average reflectivity, focused on a spot of 2 cm radius at wavelengths between 290-650 nm, is around 80\%.

The light structure allows the pointing direction of the telescope to change very fast. But due to changing gravitational loads the shape of the reflective surface gets slightly modified depending on the pointing direction. The Active Mirror System ensures optimal focusing for each telescope pointing. Each mirror panel can be individually moved through two computer-controlled actuators and they are used to automatically adjust each mirror when the telescope moves.

\subsection{The camera}

The camera is a critical element to reach good gamma-ray sensitivity through differentiating gamma-ray induced events from those induce by either hadrons or NSB. In the case of the MAGIC telescopes their design \citep{MAGIC-CameraDesign} was done aiming at: low weight, capability to observe with bright skies (i.e. the moon above the horizon) and optimised sensitivity at low energies. 

Most of the mechanical components of the camera are made of aluminium for a total weight of 600 kg. The physical dimensions of the camera are 1.462 m diameter and 0.81 m in thickness. The camera is sealed by a 3 mm thick Plexiglas window with 94\% transmission above 340~nm and still of 80\% transmission at 300~nm to be protected from the external environment. Therefore a cooling system that is able to stabilise the temperature of the electronics inside the camera despite the heat produced by itself needs to be in place. Two cooling plates with pipes through which a cooling liquid passes provide a temperature stability inside the camera at the level of $\pm$1 degree for external temperatures between -10 and 30 degrees.

The cameras of the MAGIC telescopes have a sensitive surface of 3.5 degrees diameter FoV with a trigger region of 2.5 degrees diameter FoV. These cameras differ mainly from the camera initially mounted in the first MAGIC telescope \citep{Cortina:2004qt} on the amount and size of the pixels. In addition the internal design was improved, significantly facilitating the maintenance activities. The current cameras are equipped uniformly with 1039 pixels of 0.1 degrees diameter. The original camera had an inner hexagonal area composed of 397 pixels with 0.1 degrees diameter FoV surrounded by 180 0.2 degrees diameter FoV pixels. The pixels have an hexagonal shape with PMTs from hamamatsu (R10408) as the photosensors, with a typical peak efficiency of $\sim$ 32 $\%$. These PMTs have only 6 dynodes, which allow them to be run at a relatively low gain (40 000) limiting their ageing under bright skies. The electrical signal from the PMTs is amplified about 25 dB with an AC coupling and then transformed to optical signal by means of vertical cavity surface emitting lasers (VCSELs)\citep{MAGIC-Upgrade}. This allows transmitting the signal from the camera to the Control House housing the trigger and readout electronics with minimal losses. Having only the conditioning electronics on board of the camera made possible to reduce its weight and facilitated partial upgrades. Transmitting the signal through optical fibers provides good integrity.
The signals coming from a single photoelectron measured with the acquisition system have a pulse width of 2.5ns.

\subsection{Receivers and the trigger systems}

The optical signals from the camera are transmitted through 160 m long optical fibers to the Counting House, where they are converted back to electrical signals. 
These are split into two branches, the so-called digital and analog branch. 
The former is used for the standard MAGIC trigger decision system. 
The latter is mainly used for the data acquisition (DAQ) system, however it is also exploited in the alternative trigger (Sum-Trigger-II).  

The standard trigger of MAGIC consists of three steps, historically referred to as L0, L1 and L3\footnote{Additional L2 trigger step (topological trigger) was also studied, however it was never used in regular MAGIC operations.}.
Contrary to other IACTs, not the whole MAGIC camera participate in the trigger, but only the inner 53\% of the pixels. 
This is a partial remnant of the limitations of the original MAGIC design, that was focused on observations of point sources located at the camera centre.

The L0 trigger is implemented as part of the receiver boards. 
It is based on programmable discriminators: the threshold value as well as the time delay and width of the output signal is configurable for each pixel separately. 
Its resulting rate is being read at a frequency of 1\,Hz \citep{MAGIC-Upgrade}. 
This design allows for a quick response to changing conditions (such as stars passing through a FoV of a given pixel) in order to avoid saturation. 
Contrary to the threshold values, the delays do not vary during the observations, however the possibility to set them to a given value is essential to compensate the delays accumulated over the chain (PMT transit time, optical fibers, differences in electronic paths, etc.). 

The L0 outputs are being fed to a digital filter (L1) composed of 19 partially overlapping cells of 36 pixels each. 
Inside the macrocells close-compact next-neighbour (NN) logic is implemented with different multiplicities \citep{MAGIC-Upgrade}. 
In the standard stereoscopic observations 3NN logic is used. 
Alternatively, occasional monoscopic observations are using 4NN topology. 
Such monoscopic observations are performed either in the case of a temporary failure of one of the telescopes or during special runs (one hour every moon period) aimed at gathering large statistics of muon events used for optical performance monitoring. 
In such a case, the L1 is the final trigger, resulting in a rate of the order of $\sim0.5$\,kHz transmitted to the DAQ.
In contrary, for stereoscopic observations the optimal L1 trigger rate is much larger, $\sim10$\,kHz, however it is further limited by two-telescopes coincidence condition.

The output signals of L1 triggers are stretched to a gate length of 100\,ns and fed into the common L3 trigger system. 
In order to correct for the geometry of the two telescopes and the shower axis, the L1 signals are delayed according to tabulated azimuth and zenith pointing direction,
The 100\,ns gate (corresponding to maximum absolute difference of the arrival times from both telescopes of 200\,ns) is enough to ensure basically 100\% efficiency of the stereoscopic trigger for showers. 
The typical stereo rate is $\sim0.3$\,kHz.

The MAGIC telescopes are equipped also with an alternative, low-energy trigger, the so-called Sum-Trigger-II \citep{2021ITNS...68.1473D}.
Its concept is based on the fact that the smallest showers may have signals distributed over many pixels, which are not possible to fish out on pixel-by-pixel basis from the NSB noise.
However while the signals would stack up linearly, the NSB noise will only grow with a square root of the number of pixels. 
Therefore the analogue information from patches of 19 pixels are summed up into one pulse, which is then compared with a threshold. 
The size of those patches was optimized to cover an approximate angular extend of $\sim$80\,GeV showers observed at low zenith angle.
In order to avoid the output pulse being dominated by a single afterpulse-driven noise signal, the signals are clipped before summation. 
The optimal performance of the trigger through the summation procedure requires also careful adjustment of the timing of individual channels, more precise than in the case of the standard L1 trigger. 
The output of the Sum-Trigger-II in each telescope is than combined into a stereoscopic trigger similarly to the case of the standard trigger. 

The performance of the Sum-Trigger-II was shown to be superior to the standard trigger in the energy range below 100\,GeV (for low zenith distance observations) \cite{2021ITNS...68.1473D}.
At 20\,GeV the improvement of the trigger collection area is an order of magnitude. 
At higher energies the performance is nearly the same (slightly worse in the case of Sum-Trigger-II due to slightly smaller effective trigger geometrical area).
A caveat of the Sum-Trigger-II observations is twice larger DAQ rate, that both slows down the fast processing of the data and increases data storage costs. 
Therefore the trigger is selected on a particular science case.
While most of the observations are performed with the standard trigger, Sum-Trigger-II is used for targets that can heavily exploit lower energy threshold, such as pulsars and distant AGNs.

\subsection{The readout}

The sampling electronics digitises the pulse signals and the DAQ program takes care of the final event building and raw data storage. The digitisation concept of the current MAGIC cameras was introduced with MAGIC-II in 2008 \citep{MAGIC-DRS2, MAGIC-DRS4}. The sampling and acquisition system is built around a modular architecture based on the Pulsar board, a 9U VME multipurpose motherboard card \citep{MAGIC-PulsarBoard}. The core of the sampling electronics is the DRS4 waveform digitizer\cite{DRSWeb}, a switching capacitor array of 1024 units with configurable sampling frequency (from 0.7 to 5 GS/s). The sampling and acquisition system for one MAGIC telescope is composed of 12 Pulsar boards, for a total of 1152 readout channels digitised synchronously. The capacitor signals are read out using an internal shift register clocked at 32 MHz using a 14-bit nominal resolution standard analogue to digital converter. The sampling speed for DRS4 chips is set to 1.64 GS/s. Sampling frequency at the level of 1-2 GS/s helps to improve the sensitivity to gamma rays allowing to better reconstruct the light pulse reaching the camera photo-sensors from them. On the other hand, the faster the sampling, the deeper the buffer in the readout systems should be. The buffer should be large enough to check for coincidence between the two MAGIC telescopes when pointing to high zenith angles, since the light front will need to travel the 80 m separating them. The 1.64 GS/s keeps the optimal performance while allowing to cover basically the full sky. The total dead time is 27 $\mu$s for each event being dominated by the DRS4 readout time.

The Data AcQuisition (DAQ)\citep{MAGIC-Upgrade, MAGIC-DRS2} program is a multi-threaded C++ program that takes care of: reading data samples from sampling electronics, building the actual events by merging the different channels and storing the data as raw files in a dedicated RAID disk system. The DAQ is also able to identify corrupted data as well as to perform basic calibrations and low-level analysis. The three main threads are: reading, analysing and writing. The computer where the DAQ runs has a RAM memory shared with PCI cards installed in it. Those PCI cards fill the RAM memory with the data coming from the sampling electronics and the reading thread proceeds to build the events whenever new data fills the shared memory. Once the event is built, the analysing thread takes care of performing low level calibrations and a simple signal intensity and arrival time reconstruction. This result of that real time analysis allows the MAGIC telescopes to have an online analysis software that provides detection plots and sky maps while the observations are being performed. Finally, the writing thread stores the already analysed events in a dedicated RAID disk. The DAQ can handle a sustained rate of 1 kHz and the rate reaching it is actually artificially limited to 1 kHz to avoid saturation problems.

\subsection{The Data Centre}

The MAGIC Telescopes are acquiring data during night. The available time for observations ranges between 6 and 11 hours per night, depending on the season of the year. Every moon cycle the observations are stopped during 3 or 4 days, because of the effect of the full moon light. Weather conditions may also prevent data taking. All together leads to about 200 TB per year of stored raw data. 

Once the observation night finishes the On Site Analysis and the DataCheck software is run over the data stored by the DAQ software on disk that night. The former produces already calibrated and higher level data and the latter analyses the data to validate its quality. Both raw data and analysis products from the On Site Analysis and DataCheck are transferred from the observatory site to the data centre located at the Port d’Informació Científica (PIC). This has bee done through a custom data transfer software developed at PIC that run on top of the LCG File Transfer Service for many years. But a systems based on RUCIO\cite{rucio2019} will start operating beginning 2023.

The MAGIC data centre at PIC provides services to both the members of the MAGIC collaboration and the full scientific community. MAGIC members can access MAGIC data from raw format to the  highest analysis products. This includes the needed computing resources for fast recalibration of raw data on demand. An Observations Database, the Software Repository and an internal Wiki to discuss ongoing data analysis together with a complete User Support platform are services also provided to the full collaboration by the Data Centre. Finally, the data centre provides to the scientific community public MAGIC results through a Virtual Observatory server, and the release of public MAGIC results in FITS format.

The Data centre includes a Data Management and Preservation Plan (DMPP), which includes a Long Term Legacy Release (LTLR) protocol for calibrated data. The LTLR implies that the raw data are kept for 5 years and afterwards only specifically relevant samples. The calibrated data is kept on disk for two years but all older data is easily recoverable from tapes. The DMPP allows both to ensure the data legacy of the MAGIC collaboration and to keep the cost of the data centre at reasonable levels.

\subsection{From mono to Stereo}

The most important upgrade of the MAGIC telescopes was the construction of MAGIC-II that concluded in 2008. 
There is a qualitative difference between observations with a single telescope and with a stereoscopic system of IACTs.
A single telescopes sees a 2-dimensional angular image of the atmospheric shower. 
Therefore a low impact event that developed deep into the atmosphere can be confused with an event at higher impact distance, but with the emission seen from higher levels of the atmosphere.
This is particularly problematic in the case of single muon events passing at a few tens of meters from the telescope and mimicking gamma-like images. 

In contrary, if two or more telescopes are located within the light pool ($\sim100$\,m) they are able to observe simultaneously the same shower but from a different angle. 
This allows 3-dimensional stereoscopic reconstruction of the shower (see e.g. 
\cite{1996APh.....5..119K, 1999APh....12..135H}) the simple calculation of the basic parameters of the shower (impact parameter, height of the shower maximum, source direction) directly from the observed geometry of images and telescope positions. 
Stereoscopic observations are a great way of significantly reducing the single muon-induced gamma-like background (see e.g. \cite{MAGIC:2011kvf}).
First, single muons require to hit a favourable impact points to be able to trigger both telescopes. 
Second, they are reconstructed to have its maximum of observed light at about 2\,km above the telescopes, which is well separable from the low-energy gamma-ray showers that are typically $\sim10$\,km above the telescope level.

Due to the above mentioned effect, and the general better reconstruction of the shower, stereoscopic observations provide high background reduction and better sensitivity. 
It is most pronounced at the lowest energies that are limited by the systematic background stability condition creating a unbreakable sensitivity limit, no matter on how big the exposure is.

The construction of MAGIC-II hence provided a factor 2 improvement in the sensitivity (see Fig.~\ref{MAGIC-HistoryPeformance}).
Such improvement was achieved even if with a single MAGIC telescope independent information of the impact parameter could be obtained using timing parameters \cite{MAGIC-ImageCleaning} thanks to the isochronous telescope design.

\subsection{MAGIC upgrades}

The design choice to transfer the signal from the photo-sensors out of the camera analogically and continuously allowed several upgrades of both the trigger and readout system. Although the major upgrade in the years 2011-2012 included the exchange of the oldest camera.

In February 2007 the initial readout system based on a Flash Analog-to-Digital Converter (FADC) running at 300 MS/s with two overlapping gains of 8 bit resolution \cite{Cortina:2002qh} was replaced by a low cost 2GS/s readout system (MUX). This allowed the MAGIC telescopes to exploit the fast flashes of Cherenkov light from the extensive air showers. The new system used the Fiber-Optic Multiplexing technique that allowed the sampling of the signal coming from 16 Photomultiplier tubes with a single 2 GS/s FADC \cite{Goebel:2007cb}. In addition the readout system for the second telescope that become operative in 2009, used a different solution \cite{MAGIC-DRS2} based on a switched capacitor array implemented in an application-specific integrated circuit (ASIC) (DRS2 chip, earlier version of the currently used DRS4 solution).

\begin{figure}
    \centering
    \includegraphics[width=0.455\textwidth]{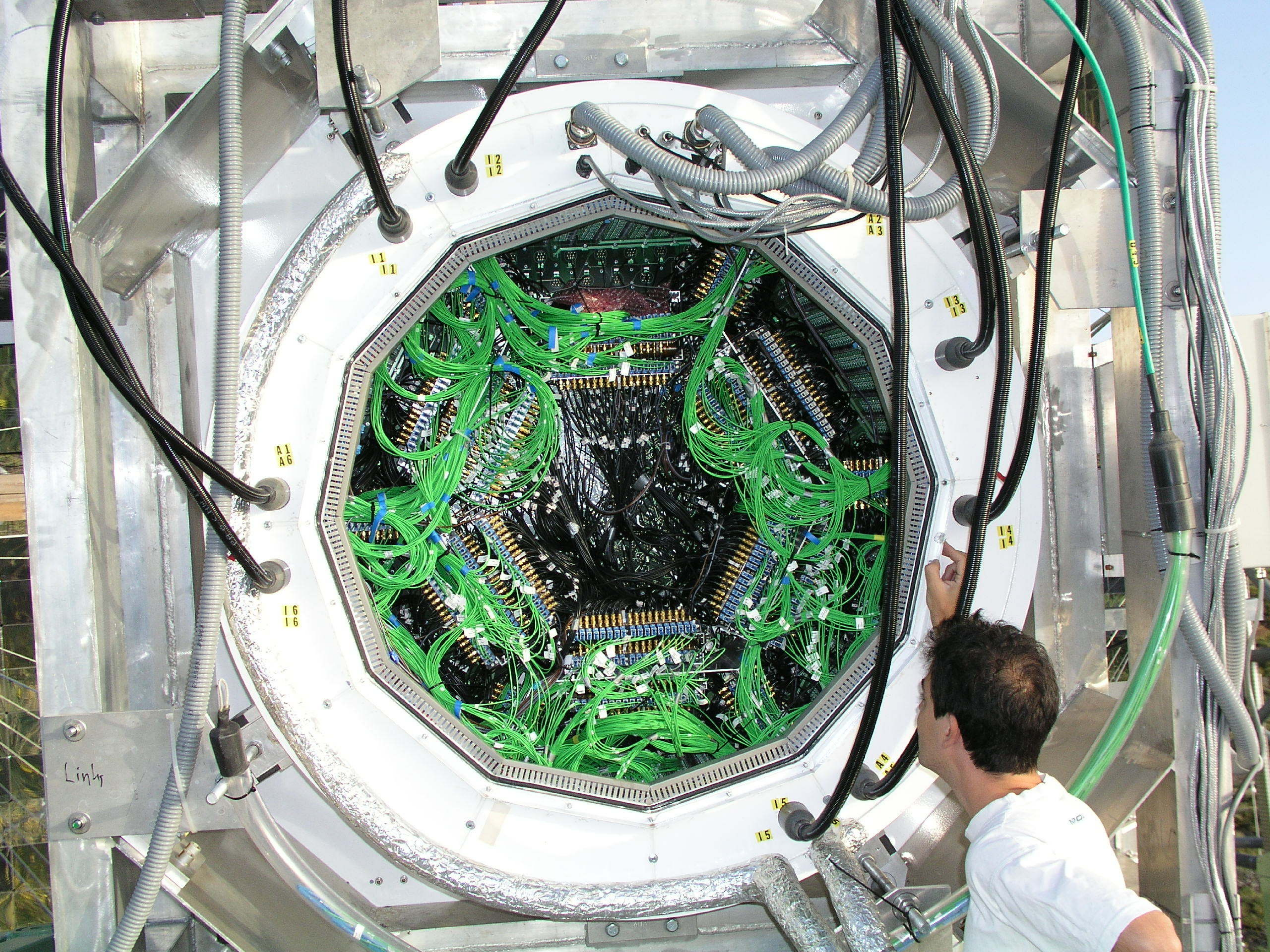}
    \includegraphics[width=0.51\textwidth]{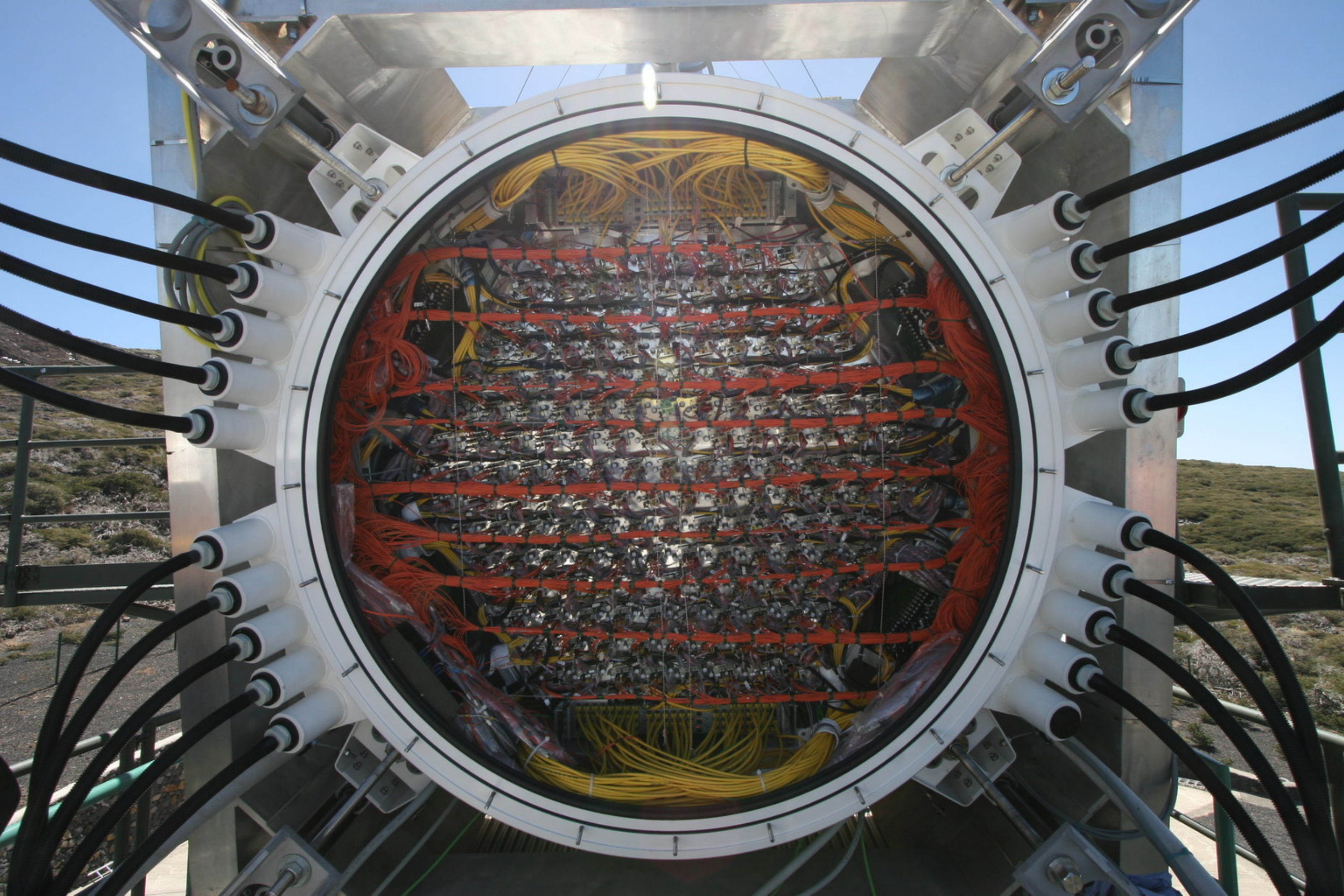}
    \caption{Comparison of the back side (after removing  protection) of the original MAGIC-I camera (left) and MAGIC-II camera (right). Photo credits: M. Garczarczyk}
    \label{fig:camera_back}
\end{figure}

The major upgrade of the MAGIC telescopes \cite{MAGIC-Upgrade} that took place in the years 2011-2012 affected the camera, the receivers and the trigger of the first telescope as well as the readout of both telescopes. The changes that affected only the first telescope were mainly aiming to unify the hardware for both telescopes. The change in the readout was mainly driven by an improved version of the DRS chip, the DRS4. This major upgrade led both to easier operation and maintenance (see Fig.~\ref{fig:camera_back}) of the system as well as improved performance \cite{MAGIC:2014zas}

\section{The MAGIC performance}


Transient events are basically happening at an unknown time and position. Hence, telescopes with a small field of view like MAGIC usually need to be repointed after receiving an alert from large field of view instruments. The faster the repointing is done, the earlier phases of the transient events can be studied. Both the design of the telescope and the Automatic Alert System put in place are instrumental for a fast repositioning to GRB or other fast transient alerts. 

The satellite born instruments have difficulties to reach good sensitivity to high energy gamma rays because the signal produced by them saturates the detector and because of their limited effective area. On the other hand, ground based gamma-ray detectors have limitations at low energy because the signal produced by those gamma rays become dimmer and eventually cannot be extracted from the NSB noise. The lower the energy of the gamma ray reaching the atmosphere is, the lower the amount of produced Cherenkov light is. Both a larger mirror surface and a more efficient camera help to collect more Cherenkov light induced by a single gamma ray and, hence, increase the sensitivity to lower energies. In addition, the relatively high location of the MAGIC telescopes (2.2 km a.s.l.) also improve sensitivity to the lowest energies.

\begin{figure}
    \centering
    \includegraphics[width=0.9\textwidth]{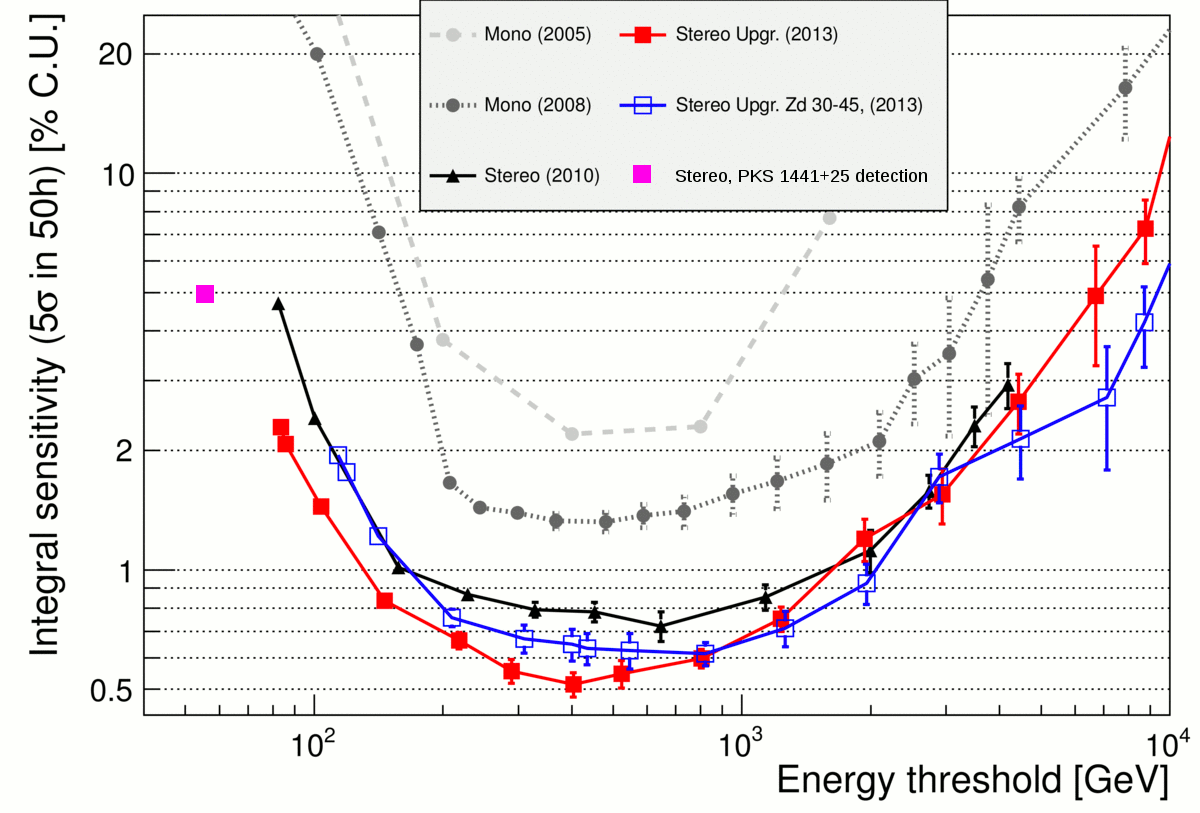}
    \caption{Evolution of the integral sensitivity of the MAGIC telescopes. Gray circles: sensitivity of the MAGIC-I single telescope with the 300 MS/s FADC (light gray, long dashed) and MUX readouts (dark gray, short dashed). Black triangles: stereo before the upgrade. Squares: stereo after the upgrade: zenith angle below 30 degree (red, filled), 30-45~degree (blue, empty) \cite{MAGIC:2014zas}. The added pink point shows the integral sensitivity above 55 GeV obtained with the PKS1441+25 detection sample \cite{2015ApJ...815L..23A}.}
    \label{MAGIC-HistoryPeformance}
\end{figure}

The integral sensitivity of the MAGIC telescopes in their different configurations since the first telescope was built, until the major upgrade \cite{MAGIC-Upgrade, MAGIC:2014zas} that took place in the years 2011-2012 is shown in Figure~\ref{MAGIC-HistoryPeformance}. The improvement over time depends on the energy range of interest. It reached a factor 10 for energies as low as 100 GeV and a factor 4 around 400 GeV. Such an improvement was possible by the combination of hardware upgrades and the implementation of new software analysis methods.

\subsection{Sensitivity}

The sensitivity indicates the capability to detect a source with a given flux, provided that it is observed for a particular amount of time. Both integral and differential sensitivities are commonly used. The former refers to a flux above a given energy while the latter refers to a flux in a narrow energy range. 

In MAGIC, the integral sensitivity (sometimes referred as sensitivity) was historically defined as the integrated flux of a source above a given energy for which the number of excess events ($N_{excess}$) divided by the square root of the number of background events ($N_{bkg}$) was equal to 5 after 50 hours of effective observation time. In addition, we require that $N_{excess}$ $>$ 10 and $N_{excess}$ $>$ 0.05 $N_{bkg}$. The corresponding energy threshold is computed as the peak of the true energy distribution of a Monte Carlo sample with a Crab-like spectrum. This is the receipt to compute all the sensitivities shown in Figure~\ref{MAGIC-HistoryPeformance}. The result of $N_{excess}$ divided by the square root of $N_{bkg}$ gives a rough estimation of the probability to get that signal from a random fluctuation of the background in units of standard deviations. A better estimation of this probability is provided in \cite{1983ApJ...272..317L}. In particular, formula 17 is the standard method in the VHE gamma-ray astronomy for the calculation of the significance. The MAGIC collaboration has also adopted this definition and the sensitivity of MAGIC computed according to that with 5 background regions is (0.67$\pm$0.04)\%~C.U. above 290 GeV for 50 hours of observation. Details of the computation as well as the sensitivity with several other assumptions are provided in \cite{MAGIC:2014zas}. Even while 50 GeV is usually below the analysis threshold of the MAGIC telescopes (for a standard stereo trigger) for a Crab-like spectrum, it is still possible to perform scientific observations with the MAGIC telescopes at those energies. Moreover, for sources with a steep spectrum the peak of the true energy distribution shifts to lower values. For example, a strong 25 sigma detection of the $z=0.939$ flat-spectrum radio quasar PKS 1441+25 \cite{2015ApJ...815L..23A} corresponds to 5\%~Crab flux sensitivity above 55 GeV for 50 hours of observations.

\begin{figure}
    \centering
    \includegraphics[width=0.9\textwidth]{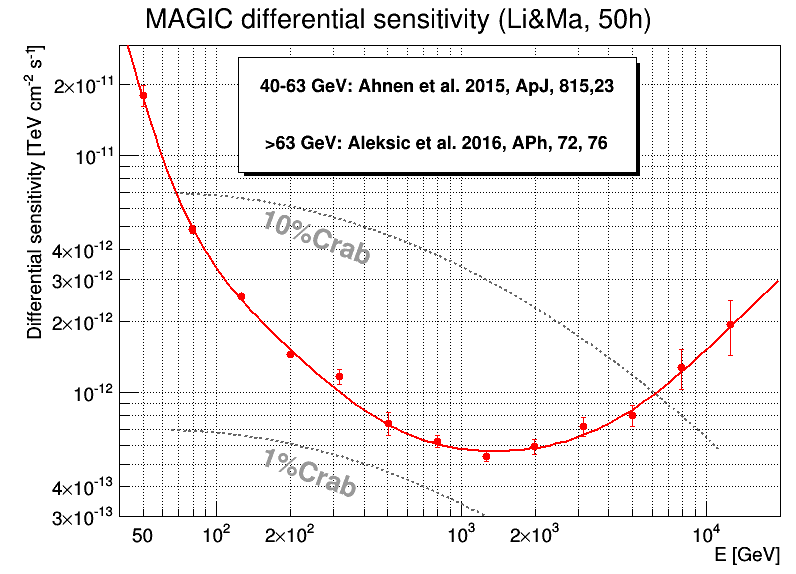}
    \caption{Differential sensitivity of MAGIC in TeV cm$^{-2}$ s $^{-1}$ .}
    \label{DifferentialSenstivityAbsolute}
\end{figure}

The differential sensitivity allows better assessment of the performance for sources with an arbitrary spectral shape. The differential sensitivity shown in Figure~\ref{DifferentialSenstivityAbsolute} is computed for 50 hours of observations requiring 5 sigma signal as given by formula 17 in \cite{1983ApJ...272..317L} assuming 3 background regions and considering 5 bins per decade of energy. It is also required that the number of detected gamma-rays is larger than 10 and than 5\% of the background. Points above 63 GeV are obtained with Crab Nebula data \cite{MAGIC:2014zas}, the 40-63 GeV point is obtained with PKS1441+25 data \cite{2015ApJ...815L..23A}.

The sensitivities of MAGIC, as well as for most of the instruments sensitive to VHE gamma-rays,  are given either in Crab units or in energy flux (TeV cm$^{-2}$ s$^{-1}$). Both magnitudes are connected through the flux measured for the Crab Nebula \cite{MAGIC:2014zas}, which is the standard candle used in VHE gamma-ray astronomy. If the sensitivity is measured through observations of the Crab Nebula, the former is a direct measurement while the latter relies on the capacity of the instrument to recover the actual flux of the Crab Nebula. If the sensitivity does not come from observations of the Crab Nebula, both magnitudes are equivalent for a given instrument and the latter provides an easier comparison to other energy ranges.

Most of the ground-based instruments sensitive to gamma rays report sensitivities for 50 hours of observations using formula 17 in \cite{1983ApJ...272..317L} and requiring $N_{excess}>$10 and $N_{excess}>$0.05 $N_{bkg}$. There are still quite some other choices like number of background regions or using Crab Nebula data that may differ. Hence, special care needs to be put on possible comparisons.

\subsection{Angular and energy resolution}

Two other commonly quoted performance parameters of Cherenkov telescopes are angular and energy resolution. 
The angular resolution is a measure of the spread of the reconstructed arrival directions from a point source.
Most commonly it is quantified as a certain (e.g. 68\%) containment region of the signal. 
Alternatively, the gamma-ray signal can be fit with a particular shape, most commonly 2-dimensional Gaussian or a combination of two such Gaussians, but other possibilities, such as the King's function \cite{2018APh....98....1D} can be used as well. 
The angular resolution (expressed as 68\% containment radius) of the MAGIC telescopes is about $0.16^\circ$ at 100\,GeV and drops to $\sim 0.07^\circ$ at the TeV energies. 
Noteworthy, the angular resolution is not only important in studying morphology of extended sources, but also has an effect on the sensitivity of point-like source analysis in terms of limiting the size of the region from which background needs to be integrated.

\subsection{Systematic uncertainties}

The IACT technique relies on the knowledge of the development of extensive air showers in the atmosphere, which is used as a calorimeter and thus effectively a part of the instrument. Therefore, to reach particular accuracy one needs not only precise knowledge of hardware built (photosensors efficiency, mirror reflectivity, telescope pointing accuracy) but also of the atmosphere and the development of extensive air showers in it. In addition, test beams in controlled conditions are not possible due to the nature of IACT technique. Alternative solutions are used like standard astrophysical candles, muons, stars in the FoV or artificial light sources. But all of them rely on assumptions that introduce further systematic uncertainties. There are many individual factors which are only known with a limited precision, thus contributing to systematic errors. Some of them may vary from one night to another or even at shorter time scales.

The systematic uncertainties induced by all different factors have been studied \cite{MAGIC:2011kvf,MAGIC:2014zas} and the uncertainties induced in the main observables are summarised in Table \ref{table:systematics}.

\begin{table}[h!]
\centering
\begin{tabular}{||l|c||} 
 \hline
 Observable & Systematic Uncertainty  \\ 
 \hline\hline
 Pointing Accuracy & $\lesssim$ 0.02 degrees \\ 
 Absolute Energy Scale & $<$ 15 \%  \\
 Flux normalisation, below 100 GeV & 18 \% \\
 Flux normalisation, 100 GeV - 1 TeV & 11 \% \\
 Flux normalisation, above 1 TeV & 16 \% \\
 Spectral slope, strong source & $\pm$ 0.15 \\
 Run to run integral flux for Crab-like spectrum   & $\sim$ 11 \% \\
 \hline
\end{tabular}
\caption{Systematic uncertainties for the main physic high level magnitudes obtained from MAGIC observations.}
\label{table:systematics}
\end{table}

\subsection{Special Observation conditions }

The sensitivity as well as the angular and energy resolution are basically enough to characterise the capabilities of the instrument to extract information from a given observation. Still, they may change depending on the observation conditions: zenith range of the observations, off axis observations, alternative trigger observations and observations under different levels of moon light.

The performance of the MAGIC telescopes changes with the zenith angle at which the observations are conducted. The larger the zenith angle, the larger minimum energy to which the telescope is sensitive and the larger its collection area at high energies \cite{MAGIC:2014zas, MAGIC-HighZenith, MAGIC:2020xry}. Hence, larger zenith angle means not having access to the lowest energies but improving the sensitivity at high energies. The effect becomes larger for very high zenith angles increasing the minimum energy to which the MAGIC telescopes are sensitive for observations at 80 degrees to 10 TeV but increasing the collection area for energies above 70 TeV by one order of magnitude \cite{MAGIC:2020xry}. This allowed to extend the measured spectrum of the Crab Nebula with MAGIC up to 100 TeV \cite{MAGIC-Collaboration:2020hl}. Those very large zenith observations require careful evaluation of the atmospheric transmission since it is much more severe and could increase the systematic uncertainties.

The observations with the MAGIC telescopes are usually carried out in the wobble mode \cite{Fomin:1994aj} with the source at 0.4 degrees from the camera centre. Still, extended sources, follow up of alerts with large position uncertainty or serendipitous sources in the FoV \cite{2010ATel.2510....1M} may lead to a region of interest at a different distance from the camera centre. The off-axis sensitivity of the MAGIC telescopes \cite{MAGIC:2014zas} stays nearly constant up to 0.4 degrees from the camera centre, slowly decreasing afterwards, and reaching a factor 2 worsening at 1.4 degrees. For larger distances the sensitivity decreases faster. The effect on the energy and angular resolution for off-axis observations is not significant up to 1.4 degrees from the camera centre.

As indicated in the hardware description of the MAGIC telescopes a lower energy trigger named Sum-Trigger-II \cite{2021ITNS...68.1473D} is available. For observations done with the Sum-Trigger-II, the collection area at the lowest energies increase being a factor 10 larger for 20 GeV gamma rays than that of the digital trigger.

The best performance of the MAGIC telescopes is achieved under dark conditions. On the other hand, extending the observations into less favourable light conditions allows to increase the duty cycle of the instrument. Being able to extend the observation time does not only provide the possibility to observe more targets in a year but also optimises the capability to follow transient events and flaring sources. The MAGIC telescopes are able to perform observations even if the sky is 30 times brighter than after dusk and in the absence of the Moon (dark condition). This means that observations can be carried out even during the full moon period. Although for brightness above 15 times that of the dark conditions, UV filters need to be installed manually in front of the camera entrance window. The main effect of moonlight is an increase in the analysis energy threshold and in the systematic uncertainties on the flux normalization. No worsening of the angular resolution was found and the degradation of the sensitivity is below 10\% even for light condition 5-8 times brighter than dark sky  \cite{MAGIC:2017zph}.

Finally, it is worth mentioning that despite most of the observations being scheduled before the data taking starts, there are occasions in which the MAGIC telescopes are re-pointed based on alerts provided by other instruments, for instance GRB alerts. In this case the capability of the MAGIC telescopes to start data taking at nominal performance is instrumental. The maximum repositioning speed is more than 7 degrees per second, meaning the MAGIC telescopes can be pointed to any point on the observable sky in less than 25 seconds. Once the telescopes are tracking a position in the sky, configuring them to optimal stable performance takes 10 to 15 seconds.

\section{The MAGIC scientific achievements}

In this section we provide an overview of the scientific results of MAGIC for various classes of objects and scientific topics. 

\subsection{Pulsars}

A pulsar is a highly magnetised, rapidly rotating neutron star that emits beams of electromagnetic emission, which is seen as pulsed from Earth. The spectra observed from pulsars with gamma ray-satellites \cite{Abdo_2013} show a cut-off at a few GeV. But ground based telescopes can explore higher energies providing insights on the acceleration mechanisms in place. 

The first pulsar detected at VHE was the Crab pulsar. MAGIC detected pulsation above 25 GeV \cite{2008Sci...322.1221A}. Those energies were already challenging one of the most accepted models for pulsar emission, the polar cap scenario \cite{baring2004high}. Further observations of the Crab pulsar revealed electromagnetic emission extending to even higher energies and excluded the existence of a cut-off in the spectrum at even hundreds of GeV. The MAGIC telescopes detected gamma rays reaching 1.5~TeV \cite{2016A&A...585A.133A}. Both the extension of the emission to those energies and the discovery  of a bridge emission above 50 GeV between the two main pulses \cite{2014A&A...565L..12A} made necessary  a deep revision of the models to explain VHE gamma-ray emission in pulsars.

Very few pulsars have been detected at VHE. Hence, it is unclear if the observed extension of the spectrum to TeV energies for the Crab pulsar is a common characteristic or not. Still, MAGIC also detected the emission of pulsed gamma ray emission from the Geminga pulsar (PSR J0633+1746) between 15 GeV and 75 GeV \cite{2020A&A...643L..14M}. The joint analysis of MAGIC and \textit{Fermi}-LAT data excludes even a sub-exponential cut-off in the explored energy range.

\subsection{Binary systems}

A large fraction of the stars in our galaxy are in binary or higher order systems. Still only few of them have been found to emit in the gamma-ray domain~\cite{Abdollahi_2020}. Those for which the gamma-ray emission is the bulk of their non-thermal emission are called gamma-ray binaries. All of them are composed of a massive star and compact object, but the nature of the latter, which has implications in the scenarios to explain the VHE gamma-ray emission, is unknown for most of them. 

MAGIC has been searching to unveil the nature of gamma-ray binaries. MAGIC first discovered TeV emission from LS I+61º303 \cite{2006Sci...312.1771A} and performed many observational campaigns leading to the discovery of the long-term variability in the VHE domain \cite{MAGIC:2016oil} and correlation between VHE gamma rays and X-rays \cite{Anderhub_2009}. MAGIC (together with VERITAS) also discovered TeV emission from  the unique system PSR J2032+4127/MT91 213 during the periastron passage, which takes place once every 50 years \cite{VERITAS:2018qgg}. A long term study of HESS J0632 has also been performed together with H.E.S.S. and VERITAS \cite{VERITAS:2021ogb}.
MAGIC has also carried out  observations on other binary systems that could potentially emit VHE gamma rays. This includes massive microquasars such as MWC 656, Cygnus X-1, Cygnus X-3  or SS433, as well as low-mass X-ray binaries like GRS 1915+105, Sco X-1 or V404 Cygni. All these observations led to upper limits.

Both the detections and deeper studies as well as the upper limits allowed us to narrow down possible scenarios for the mechanism in place in the individual sources. Still, the detection of more gamma-ray binaries and their thorough study are needed to be able to have a well motivated model of the mechanism in place in gamma-ray binaries that allow the production of gamma rays.

MAGIC has been also conducting for over a decade a special follow up program of novae \cite{2015A&A...582A..67A}.
Novae are cataclysmic variable binary systems, in which transfer of matter from the donor star to a White Dwarf companion eventually leads to thermonuclear explosion.
A few novae has been followed by MAGIC in their first few days after the optical detection. The observations were used to constrain a scenario of simultaneous acceleration of protons and electrons on the nova shock \cite{2015A&A...582A..67A}.
In 2021 the first nova, RS Oph, has been detected independently by H.E.S.S. \cite{2022Sci...376...77A} and MAGIC \cite{2022NatAs...6..689A}.
MAGIC observations, and their interpretation, for the first time showed strong evidence of hadronic origin of the gamma-ray emission from novae. 

\subsection{Gamma Ray Bursts}

A special case of ToO alerts are the so-called Gamma Ray Bursts (GRBs). 
Depending on their duration of the bulk of their emission they are classified into short ($<2$\,s) or long ($>2$\,s) GRBs.
Even in the case of the long GRBs the bulk of the emission is observed typically over a time scale not longer than a minute. 
Therefore, it is of utmost importance to start the observations as soon as possible. 
Therefore, a special Automatic Alert System was implemented in the MAGIC telescopes.
Thanks to it, it is possible to observe GRBs already a few tens of seconds after their onset (if the visibility conditions allow it). 
After 15 years of optimising this procedure and GRB analysis methods, MAGIC observations revealed emission from a long GRB 190114C, the first GRB detection claimed by IACTs \cite{2019Natur.575..455M}. 
The observations started already 50\,s from the onset of the burst, and in the first seconds the source was by far the brightest VHE source ever - reaching flux of 100 times of the Crab Nebula. 

Despite significant absorption of the emission in the 
EBL, the spectrum of GRB 190114C was measured up to TeV energies. 
This allowed to undoubtedly confirm that the gamma-ray emission of GRBs is not a simple extension of the observed at the lowest energies synchrotron emission, but a separate component. 
It can be explained by a inverse Compton emission, similarly as in the blazar models \cite{2019Natur.575..459M}. 

More recently another long GRB was detected by the MAGIC telescopes, GRB 201216C \cite{2020ATel14275....1B}. 
With the tentative redshift of 1.1 it is currently the most distant VHE gamma-ray source known.

\subsection{Monitoring of Bright AGNs}

Typically about 40\% of the MAGIC observation time is devoted to studies of AGN objects \cite{2017ICRC...35..658S}.
Those include both search for VHE gamma-ray emission from sources not yet known in this energy range (often in response to a high state at lower energies), but also monitoring of known VHE gamma-ray emitters. 
Re-observations of known sources are justified because of strong variability of AGN objects. 
Also, the monitoring observations are performed together with novel instruments operating at lower energies - providing more global view of the source emission.
In order to maximise the scientific output of those observations, the brightest objects are selected, however of a few different classes.

In the case of HBL objects the two archetypal sources are Mrk\,421 and Mrk\,501. 
Both can be relatively easily detected, even if a low emission state. 
The observations proven to be very fruitful, resulting in about 20 papers describing light curves, interband correlations (see e.g. \citealp{2021MNRAS.504.1427A}). 
The simultaneous, broadband spectra were also interpreted in short time windows, mainly with a simple 1-zone homogeneous Synchrotron-self-Compton (SSC) scenario. 
In fact, on multiple occasions showing limitations of such a simple approach, such as correlations of the model parameters \cite{2017A&A...603A..31A}.
In the case of observations during high X-ray activity, a hint of additional, narrow TeV spectral component was observed \cite{2020A&A...637A..86M}.
The observations of those two sources allowed also to search for an extended emission - in the framework of cascades deflected in the intergalactic magnetic field (IGMF) \cite{2010A&A...524A..77A}.

Another, more distant HBL commonly observed with MAGIC is PG\,1553+113 \cite{2012ApJ...748...46A}.
Due to its high redshift (uncertain, however likely $\sim0.4$) and extreme flares it is a good candidate for studies of the gamma-ray propagation through the universe \cite{2019MNRAS.486.4233A}.
The interest in gamma-ray observations of this source was amplified when quasi periodic oscillations in optical and GeV range were reported on biannual time scale \cite{2015ApJ...813L..41A}. 
This led to intensification and revision of the monitoring scheme \cite{2017AIPC.1792e0018D}. 
Nevertheless, so far the extension of the quasi periodicity into VHE gamma-ray range could not have been proven. 

In the case of FSRQ VHE gamma-ray monitoring is more problematic. 
Those sources are often located at a high redshift, such that the absorption on the way to the observer stacks up with typically softer intrinsic spectra, significantly worsening detection prospects in short time scales. 
PKS\,1510-089 is a FSRQ at moderate redshift ($z=0.36$), with a moderately high flux in VHE band. 
It was first detected by H.E.S.S. in 2009 \cite{2013A&A...554A.107H} during a high state, and then confirmed by MAGIC during another episode of enhanced emission in 2012 \cite{2014A&A...569A..46A}.
Curiously no variability of the emission was seen in H.E.S.S. or MAGIC observation.
Also the two measurements were consistent within statistical and systematic uncertainties. 
In the view of general high variability of blazars, and also strong optical and GeV variability of PKS\,1510-089 this was a puzzling result and justified further monitoring efforts. 
In 2015 MAGIC observations of a flare from this object showed that the source is indeed variable \cite{2017A&A...603A..29A}.
Next year, further monitoring of the source (triggered by the H.E.S.S. report of an enhanced activity) revealed another VHE flare, a factor of a few brighter than 2015 one.
Interestingly the decaying part of the light curve observed by MAGIC showed a sudden cessation of the emission, that is difficult to explain with the existing models of the blazar jets \cite{2021A&A...648A..23H}. 
Finally, PKS\,1510-089 is also the only FSRQ that was detected in VHE gamma rays during low GeV state \cite{2018A&A...619A.159M}.

MAGIC telescopes are also used for monitoring of radio galaxies (RG).
While the RG are observed relatively close, and thus do not require low energy threshold, their TeV fluxes are rather low. 
The RG monitoring efforts of MAGIC concentrated around M87 \cite{2020MNRAS.492.5354M}, the archetypal RG, and RGs located in the Perseus cluster (NGC 1275 and IC 310, \citealp{2014A&A...564A...5A}). 
These observations revealed occasional extreme flares, some of them with time scales of variability of minutes \cite{2014Sci...346.1080A,2018A&A...617A..91M}. 

All those deep-exposure, MWL monitoring efforts provide a valuable insight for theorists aiming to explain long term behaviour of AGNs. 
The observations  also create a MAGIC legacy data set that will outlive the MAGIC telescopes. 

\subsection{ToO Program}

With a relatively narrow FoV and only $\sim 1000$ hrs of dark-time observations available per year it is important to prioritise IACT observations to maximise scientific output. 
In general, due to small relative statistical errors,  high states of sources are able to provide us more detailed information about the emission than low states.
Alternatively, in some cases the source undergoes a major change of its state (which can be cataclysmic, such as explosion of a nova star, or less dramatic, ``mode-switching'' type). 
The VHE emission might be expected only in some specific states of the sources. 
It is thus not only important which source will be observed, but also when. 
The scheduling of MAGIC telescopes is adaptive and able to accommodate observation requests motivated by short term activity of the source. 
This so the so-called Target-of-Opportunity (ToO). 
Typically a few tens of \% of MAGIC observation time is spend on ToO. 
Obviously it makes sense to observe only variable (or newly detected) sources in the ToO scheme. 
This partially favours extragalactic sources (in particular AGN and Gamma Ray Bursts), however a few classes of galactic sources (such as microquasars, Crab flares, magnetars, novae) can be also observed in this scheme, as well as multi-messenger (neutrino, gravitational waves) alerts.  

In the first years of MAGIC, due to lack of operating GeV satellite, the main source of extragalactic ToO alerts was optical observations (in some cases also X-ray triggers were followed)  \cite{2008ICRC....3.1033M}.
This led to the detection of a few blazars in TeV energy range. 
The game-changer was the launch of the \textit{Fermi}-LAT satellite, sensitive instrument scanning the whole sky in GeV range \cite{2009ApJ...697.1071A}.
As the GeV range is very close to the energy range of MAGIC, simple extrapolation arguments can be used to evaluate detection prospects of the sources. 
Therefore \textit{Fermi}-LAT alerts become the most dominant source of ToO observations of MAGIC. 
They led to a number of discoveries of VHE gamma-ray emission, including two high redshift sources PKS1441+25\cite{2015ApJ...815L..23A} and QSO B0218+357 \cite{2016A&A...595A..98A}. 
The caveat of \textit{Fermi}-LAT alerts is that due to satellite downlink, data integration and processing time, the GeV information is usually 24-48 hrs old, which significantly hinders observations of very fast alerts.

In the case of Galactic sources ToO alerts led to the discovery of RS Oph \cite{2022NatAs...6..689A}, first nova detected in the VHE gamma-ray range.
Also enhanced emission from other binary systems was detected (see e.g. \cite{2012ApJ...754L..10A}).

\subsection{Extragalactic Background Light}

The extragalactic background light (EBL) is a cosmic diffuse radiation field mainly composed of the light emitted by stars since their formation, which covers the wavelength from ultraviolet to near infrared. In addition, a fraction of these photons is absorbed and re-emitted at longer wavelengths by the interstellar dust. The observation of gamma rays from distant extragalactic sources allows constraints on the EBL. Photons travelling through the EBL will be attenuated due to pair production. Since the cross section of the pair production is strongly peaked to the centre of mass energy of 1.8 times the mass of the produced particles, there is a specific range in the EBL energy which is “probed” by each gamma ray energy. 

The spectra of several individual sources \cite{MAGIC:2008sib, Ahnen:2016gog} allowed MAGIC to put constraints on the EBL scale assuming a model for the evolution of the EBL with redshift. A wavelength dependent constraints on that energy scale could be established by using the spectra obtained with MAGIC and {\it Fermi}-LAT from 12 blazars in the redshift range z = 0.03 -- 0.944, which minimises the bias on the model for the EBL evolution.  The global scale for the EBL ranges from 1.00 to 1.38 depending on the  model for the EBL evolution, with uncertainties usually below 0.20 \cite{2019MNRAS.486.4233A}.

\subsection{Fundamental Physics}

Since its beginning, the MAGIC Collaboration puts emphasis on fundamental physics studies. In particular searches for Dark Matter (DM) and Lorentz Invariance Violations (LIV) effects are main topics in this area. 

Both DM annihilation or its decay products could lead to observable VHE gamma rays. Those signals are expected to come from cosmic places with high concentrations of dark matter: surrounding of intermediate black holes, Galactic Centre, dwarf spheroidal satellite galaxies (dSphs) and galaxy clusters. Observing different types of possible sources allows us to reduce systematic effects related to their different characteristics. This is also valid for different individual sources of the same type. The dSphs are expected to have small gamma-ray emission from astrophysical sources and have large mass-to-light ratios, hence, they are  among the best candidates to search for dark matter annihilation with IACT. MAGIC has observed several dSphs (Segue 1, Ursa Major II, Draco, Coma Berenices) collecting about 350 hours of good quality data and combined their observations \cite{2022PDU....3500912A} and with {\it Fermi}-LAT observations \cite{MAGIC:2016xys}. The combination of all MAGIC observations led to 95\% UL on the velocity-averaged cross-section $\langle \sigma_{ann}v\rangle$ at the level of  10$^{-24}$ cm$^{3}$/s at $\sim$ 1 TeV for most of the 9 annihilation channels considered. 

Galaxy clusters are the largest known gravitationally bound structures in the Universe. They are reaching masses of 10$^{15}$ solar masses, including larger fraction of dark matter than other astrophysical sources. MAGIC took about 400 hours on the Perseus Clusters. The absence of a signal led to the conclusion that dark matter particles have a decay lifetime longer than $\sim10^{26}$s in all considered channels \cite{MAGIC:2018tuz}.

The LIV is the deformation of the Lorentz symmetry that arises from some candidate theories for Quantum Gravity. One of the effects of LIV would be the modification of the dispersion relation for photons in the vacuum, that could be parameterised with energy-dependent corrections with the Quantum Gravity Energy (E$_{QG}$) as main parameter. Hence, the time that gamma rays need to travel from the astrophysical sources to the Earth would depend on their energy. The larger the distance and the larger the energy difference are, the larger the time difference is. The MAGIC collaboration has been looking for energy dependent time delays from several types of sources: Active Galactic Nuclei \cite{Albert:2007zd}, Pulsars \cite{MAGIC:2017vah} and Gamma Ray Bursts \cite{MAGIC:2020egb}. No significant delays have been observed and limits on the E$_{QG}$ have been established.

\section{The future of MAGIC}

In this section we discuss the possible future of MAGIC, its relation to the next generation of IACTs, and possible alternative use cases of the telescopes in the CTA era. 


\subsection{CTA North being built}

On 10 October 2018, the inauguration of the first Large Size Telescope (LST1) of the Cherenkov Telescope Array (CTA) was held. The LST1 is located at $\sim$ 100~m from the MAGIC telescopes and it is the first telescope of the CTA Northern site that is being built around the location of the MAGIC telescopes. CTA is a world wide effort to built an observatory of Cherenkov telescopes to observe VHE gamma-ray with an increased sensitivity compared to the current ones, among which the MAGIC telescopes.

As of 2022, the LST1 is still the only telescope of the CTA Northern site already built and the MAGIC telescopes stay operative without any reduction of their scientific production.

In the long term, the array of CTA telescopes will provide much better sensitivity than MAGIC, thus limiting reasonable science targets for MAGIC. However the location of the LST-1 telescope so close to MAGIC allows common analysis of the same showers observed with two system, and hence providing 3-telescopes stereoscopic system.
While there is currently no hardware trigger joining the two systems, the events can be matched using their individual time stamps. 
MC studies show that such joint observations allows a considerable improvement of the sensitivity with respect to MAGIC-only or LST-1-only data  \cite{2019ICRC...36..659D}.
To allow such analysis currently a significant fraction of LST-1 observations are being performed in the so-called piggy-back mode (following with LST-1 the same source, in the same wobble position as being observed by MAGIC). 

\subsection{MAGIC data legacy}

Despite the MAGIC telescopes are still collecting data being scientifically exploited by the MAGIC collaboration, an effort is being done to make MAGIC data open to the full scientific community at least as legacy or even before. 
MAGIC data can already be written in the DL3 format \cite{Deil:2016yrp} that is being accepted by the full gamma-ray community as the common format as well as analysed with the open software gammapy \cite{CTAConsortium:2017xaq} allowing also for simple joint analysis with other instruments \cite{2019A&A...625A..10N}. Some of the data analysis inside the collaboration are already done using DL3 and gammapy and the full chain will eventually be ready to be used by non collaboration members.

\subsection{Alternative and complementary uses of MAGIC}

Once the CTAN reaches its expected performance, much better comparing to MAGIC, the current style of MAGIC operations will need to be modified. 
This does not necessarily mean that the telescopes have to be dismantled, as there is a number of complementary to CTA, and completely alternative uses of MAGIC telescopes.
While CTAN will have a factor of a few better sensitivity than currently operating IACTs, allowing to achieve the same accuracy with about an order of magnitude shorter time, it will be also limited by only 1000\,hrs of dark time available per year. 
Taking into account that CTA will allow also for studies not possible with the current telescopes \cite{2019scta.book.....C}, but requiring deep exposures, it is expected that the time overbooking for CTA projects will be large.
Therefore one of the uses of MAGIC in the CTA era is to take part of the burden from CTA with projects that do not require very high sensitivity. 
A natural candidate is the monitoring of AGN objects. 
This is already being done e.g. by FACT \cite{2017Galax...5...18T}, however due to its small mirror size it is only possible to do it for bright objects with relatively hard spectra. 
Using instead MAGIC for AGN monitoring would allow to do so for a much larger number of objection, including also FSRQs.
Similarly, Galactic gamma-ray binaries could be also monitored, or search for TeV counterpart of fast radio bursts can be performed.

While MAGIC is primarily an IACT it can be also considered as a very particular optical instrument: with a very large mirror area and extremely good time resolution, but poor angular resolution.
This opens a few possibilities for an alternative uses of the telescopes. 
The above mentioned features of MAGIC makes it an ideal instrument for intensity interferometry.
It allows optical observations of bright sources with unmatched angular resolution (sufficient to e.g. measuring angular diameters of stars).
It is being actively pursued in MAGIC, with the first results being very encouraging: \cite{2020SPIE11446E..1CA}.
High optical angular resolution can be also achieved with the asteroid occultation method, actively pursued by VERITAS \cite{2019NatAs...3..511B} and also possible for MAGIC.

Another possible application of profound importance, exploiting large mirror area and fast sampling of MAGIC is the search of an extraterrestrial intelligence in the so-called optical SETI approach. 
Namely MAGIC could be used for searches of short but powerful LASER emissions from the direction of nearby stars, preferentially similar to the Sun that could host planetary systems. 
In fact, such an application was already considered for MAGIC in its early days \cite{2005neeu.conf..307A}.
The fate of MAGIC in the CTA era is not yet decided, and other ideas, in response to the current trends in astrophysics, might be pursued as well.


\newcommand\aj{{AJ}}%
\newcommand\actaa{{Acta Astron.}}%
\newcommand\araa{{ARA\&A}}%
\newcommand\apj{{ApJ}}%
\newcommand\apjl{{ApJ}}%
\newcommand\apjs{{ApJS}}%
\newcommand\ao{{Appl.~Opt.}}%
\newcommand\apss{{Ap\&SS}}%
\newcommand\aap{{A\&A}}%
\newcommand\aapr{{A\&A~Rev.}}%
\newcommand\aaps{{A\&AS}}%
\newcommand\azh{{AZh}}%
\newcommand\baas{{BAAS}}%
\newcommand\caa{{Chinese Astron. Astrophys.}}%
\newcommand\cjaa{{Chinese J. Astron. Astrophys.}}%
\newcommand\icarus{{Icarus}}%
\newcommand\jcap{{J. Cosmology Astropart. Phys.}}%
\newcommand\jrasc{{JRASC}}%
\newcommand\memras{{MmRAS}}%
\newcommand\mnras{{MNRAS}}%
\newcommand\na{{New A}}%
\newcommand\nar{{New A Rev.}}%
\newcommand\pra{{Phys.~Rev.~A}}%
\newcommand\prb{{Phys.~Rev.~B}}%
\newcommand\prc{{Phys.~Rev.~C}}%
\newcommand\prd{{Phys.~Rev.~D}}%
\newcommand\pre{{Phys.~Rev.~E}}%
\newcommand\prl{{Phys.~Rev.~Lett.}}%
\newcommand\pasa{{PASA}}%
\newcommand\pasp{{PASP}}%
\newcommand\pasj{{PASJ}}%
\newcommand\qjras{{QJRAS}}%
\newcommand\rmxaa{{Rev. Mexicana Astron. Astrofis.}}%
\newcommand\skytel{{S\&T}}%
\newcommand\solphys{{Sol.~Phys.}}%
\newcommand\sovast{{Soviet~Ast.}}%
\newcommand\ssr{{Space~Sci.~Rev.}}%
\newcommand\zap{{ZAp}}%
\newcommand\nat{{Nature}}%
\newcommand\iaucirc{{IAU~Circ.}}%
\newcommand\aplett{{Astrophys.~Lett.}}%
\newcommand\apspr{{Astrophys.~Space~Phys.~Res.}}%
\newcommand\bain{{Bull.~Astron.~Inst.~Netherlands}}%
\newcommand\fcp{{Fund.~Cosmic~Phys.}}%
\newcommand\gca{{Geochim.~Cosmochim.~Acta}}%
\newcommand\grl{{Geophys.~Res.~Lett.}}%
\newcommand\jcp{{J.~Chem.~Phys.}}%
\newcommand\jgr{{J.~Geophys.~Res.}}%
\newcommand\jqsrt{{J.~Quant.~Spec.~Radiat.~Transf.}}%
\newcommand\memsai{{Mem.~Soc.~Astron.~Italiana}}%
\newcommand\nphysa{{Nucl.~Phys.~A}}%
\newcommand\physrep{{Phys.~Rep.}}%
\newcommand\physscr{{Phys.~Scr}}%
\newcommand\planss{{Planet.~Space~Sci.}}%
\newcommand\procspie{{Proc.~SPIE}}%

\bibliographystyle{unsrt2authabbrvpp} 

\bibliography{bibliography}

%
%
%
%
%

\end{document}